# Leveraging Bus Mobility to Enable Communications in Urban Areas

Sabrina Gaito, Dario Maggiorini, Gian Paolo Rossi
Università degli Studi di Milano
Via Comelico 39, 20135 Milano, Italy
Email: firstname.lastname@unimi.it

*Abstract*—This paper shows that the deployment of an opportunistic network on any public transportation system leads to obtain a scalable and efficient urban communication platform. We use the term Bus Switched Networks (BSNs) to indicate this urban backbone that complements the services of 3G networks and enables to meet the application level requirements for a large class of applications by ensuring high delivery ratio and acceptable delays under different conditions of packet load. We sustain these arguments by providing three contributions. The first contribution is a novel and lightweight probabilistic routing protocol for BSN which we prove to be highly effective in satisfying the loose QoS required by urban-wide delay-tolerant information services. The second contribution is the proposal of URBeS, an analysis platform that, given a specific city served by public transportation, produces real bus mobility traces and traffic analysis for any given routing protocol. The last contribution is an extensive benchmark analysis on three real cities which have been selected to explore geo and structural diversity.

## I. INTRODUCTION

Opportunistic Networks (ONs) are a special class of Disruption Tolerant Networks (DTNs) that leverage node mobility and contact opportunities in their data forwarding to recover from network partitions and to cope with node sparsity and intermittent connectivity. Their intrinsic robustness makes ONs an appealing technology in the sphere of multi-platform network backhaul designed to support the pervasive mobile services for the ever growing multitude of mobile persons and devices. In this type of heterogeneous scenario ONs can play a twofold role. On the one hand they can help complement and offload 3G networks on the other offer natural support to a large class of slightly delay-tolerant services (e.g. content upload and the dissemination of public utility services and commercial/events advertisement) that would be provided less efficiently through the cellular network.

The deployment of an Opportunistic Network on top of a Public Transportation System (PTS) has recently drawn the attention of researchers because it inherently helps to mitigate the well-known criticalities of ONs and at the same time takes advantage of some peculiar behaviors. First, buses are powered nodes whose lifetime cannot be affected by routing operations. Second, their mobility is governed by a partially deterministic schedule. Last, a PTS involves a relatively large number of buses and this ensures a pervasive coverage of the urban area. Taken all together, the above points promise a packet delivery platform which may lead to the deployment of a robust, urban-wide, infrastructure-free, and provider-less wireless network platform with loose QoS guarantees. We call this platform Bus Switched Network (BSN).

Within a multi-channel communication context, a mobile device is supposed to select the proper wireless technology according to service availability, cost, and QoS constraints. In Fig. 1 we show a simple heterogeneous architecture where a BSN is integrated with a wired Distribution Network (DN) that could complement the services of the 3G network in a urban area. The endpoints of the communications over a BSN include

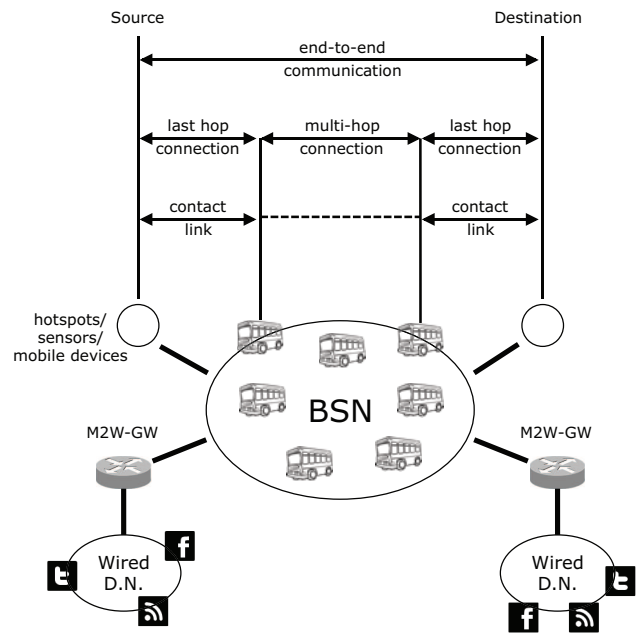

Fig. 1. Multi-platform urban architecture with a BSN and a wired distribution network.

hand-held mobile devices, environmental sensors, roadside hotspots and processes outside the BSN that can be accessed through M2W-GW (Mobile-to-Wire Gateways). We assume that the endpoints are placed somewhere along the bus routes and/or on the actual buses and that they access the BSN when entering the radio range of a bus passing by. Buses ensure the last hop connectivity to/from the endpoints; they also provide the BSN routing service that supports end-to-end communications among endpoints. Through the M2W-

GWs the packets exit the BSN and enter the wired broadband network (for instance, to upload contents to the Internet). Similarly, a reverse packet flow may transit to diffuse contents to target destinations inside the BSN.

We can imagine a large set of soft real-time mobile applications that are part of our current lifestyle and that have an urban-wide scope whose requirements are naturally satisfied by the network structure shown in Fig. 1. In general, we may organize these applications on the basis of two distinct classes. The first includes the upload of non real-time contents. In this case the sources are both sensors and mobile devices, while the destinations are usually servers outside the BSN that can be reached through a given M2W-GW. The uploaded contents include sensed data, social networks contents, plus fragments of large contents that have been fragmented to speed up the upload. In the second class we find the applications covering the download of contents (e.g., ads or information from sensing devices) on the reverse path or generated by the roadside hotspots. While the applications in the first class favor a *many-to-one* communication paradigm, those in the second adopt mainly a *one-to-any* (anycast) paradigm where the <*any*> clause may be conditioned by service-specific time/location constraints.

The end-to-end communications generated by these types of applications have a direct impact on the underlying network infrastructure. From the BSN perspective, this affects the design of both internal routing and last hop communications. Firstly, let us consider the case of internal BSN routing. Both of the abovementioned communication paradigms require packets to follow the path from <*the first encountered bus of a bus line*> to <*any bus of a bus line*>. This is a multihop path because, even in our modern cities, we cannot assume that at each bus stop there is a M2W-GW open to public service. The massive amount of messages that potentially need to be routed over BSN multihop paths imposes strong scalability constraints on the routing algorithm employed. Thus, epidemic dissemination is scarcely an option due to the level of resources consumed when scaling to urban level. Moreover, the design rationale of the routing rules has to be independent of city-specific characteristics such as city topology or PTS structure. So far, no BSN routing scheme has been designed to satisfy the needs of the described urban-wide networking conditions.

We can now show that BSNs intrinsically satisfy the requirements for last hop communication. Let us consider, as an example, buses serving as last hops for the download of contents with some location-sensitive scope of the delivery service. In this case, the geolocalized delivery is ensured inherently because last hop buses follow the same urban mobility of their users and thus any destination simply receives (or pulls) the required content locally. Interestingly, the delivery of the same service through cellular network is less natural and would be possible only under the following conditions: (*i*) the service provider establishes as many connections as the (potentially huge) number of mobile users to reach, and (*ii*) all these users are required to constantly provide their current GPS position. The first condition produces a highly inefficient use of network resources, while the second has clear weaknesses in terms of preserving user privacy (the effects of which can be reduced only by resorting to anonymization techniques).

This paper considers BSNs in the described scenario starting from recent research on the subject and focusing on the BSN routing protocol with the aim of designing a scalable solution viable for supporting urban-wide applications. We present the Op-HOP routing protocol (described in Sec. IV) that fully satisfies the abovementioned requirements. It is designed to support the specified unicast (from *one bus of a bus line* to *any bus of the destination bus line*) communications. We demonstrate that Op-HOP outperforms the classic methods based on minimizing the number of traversed hops, while at the same time avoiding the heavy resource consumption typical of epidemic-oriented or multi-copy routing, as demonstrated by the comparison of our proposed algorithm with existing multicopy schemes which expose a better behavior only at campus level. The protocol performances evaluation derive from an extensive benchmark analysis on the settings of three real cities chosen explicitly to explore geo and structural diversity (Sec. V). This section evaluates our proposal using the real PTS time schedules of Milan (Italy), Edmonton (Canada), and Chicago (IL, U.S.) as representative of all the cities reviewed during our investigation. We identify relevant city characteristics necessary to understand and predict BSNs performance by measuring city size, density of bus lines, city topology, and PTS organization. To the best of our knowledge this is the first paper addressing these issues at an urban scale. The analysis activities are based on a custom platform, named URBeS (Sec. III), which – given a specific city served by public transportation – produces realistic bus mobility traces and traffic analysis for any possible routing protocol.

## II. RELATED WORK

Despite the fact that Disruption Tolerant Networks (DTNs) were introduced only in 2003 [1], a considerable effort has been devoted by the scientific community to devising reliable routing strategies on these kinds of architectures [2], [3], [4], [5]. In particular, [5] is a multicopy forwarding scheme where probabilities are calculated on a per-bus base and adjusted with an aging policy. As special cases, in [6], [7], [8], [9], [10] the contact opportunities between nodes are exploited to build Opportunistic Networks (ONs). Among ONs, networks built on top of Public Transportation Systems (PTS) have been attracting attention in recent years.

The first contributions [11], [12], [13] focused on rural environments in developing regions where a number of villages are spread over a large territory linked by buses. The common goal of all the mentioned projects is to provide network access for elastic non real-time applications so that the local population may enjoy basic Internet services (e.g., e-mail and non-real time web browsing). In these cases the set of neighbors for every node is usually small yet does not change frequently over a span of time; failure to delivery a message is generally the result of a missed encounter and not due to an unpredictable node mobility.

Campus bus networks (e.g., [14], [15], [16]) are designed to serve students and faculties who commute between colleges or from/to nearby towns. These kinds of services are usually characterized by a relatively small number of nodes when compared to a fully fledged urban environment.

The main contribution in this direction is represented by [14], where five colleges are linked with nearby towns and to one another over an area of 150 square miles. The authors of the aforementioned paper propose a multi-copy routing algorithm, namely MaxProp, based mainly on message priorities. These priorities depend on the path likelihoods to destination nodes on the basis of historical data and other complementary mechanisms. By means of simulation MaxProp is shown to outperform oracle-based protocols hinged on knowledge of deterministic meetings between peers. This research was been extended in [15], where inter-contact time distribution were analyzed both at bus and line levels. A generative model for inter-contact times based on real traces was proposed in order to generate synthetic traces to drive simulations on routing protocols performance. On the same bus network, a system of throwbox nodes [17] was deployed to enhance the capacity of the DTN. The last case, [16], model the routing system as an allocation problem and try to optimize a specific routing metric such as worst-case delivery delay or the fraction of packets that are delivered within a deadline

Scaling up in terms of number of nodes, we find urban environments where a considerable number of lines is densely deployed to enable people to commute inside a city. Bus networks in urban environment (e.g., [18], [19], [20], [21]) are usually characterized by many contact opportunities and frequent contacts.

In [18] the authors propose a commercial application called *Ad Hoc City*. Based on a multi-tier wireless ad-hoc network routing architecture it provides elastic Internet access by means of Access Points which are responsible for a geographical area. The proposed system targets general-purpose wide area communication. Messages from mobile devices are carried to the AP and back using an ad-hoc backbone that exploits buses. The authors verified the validity of the proposed approach against real movement traces by *King County Metro bus system* in Seattle, WA. Using the same real data as for [18], the authors of [19] propose a cluster-based routing algorithm for intra-city message delivery. In [19] an efficient large-scale clustering methodology is devised: nodes are clustered based on the basis of encounter frequency while multi-copy forwarding takes place between members of the same cluster hosting the destination node. To lessen the overhead effect of having multiple copies in the network, the authors of [20] model every forwarding as an optimal stopping rule problem. The contribution from [21] uses data from the public transportation system of Shanghai to test the performance of a single-copy forwarding mechanism. This is a probabilistic routing strategy where probabilities are related to intra-contact times as in [22].

## III. EXPERIMENTAL PLATFORM

URBeS (Urban Routing Backbone Simulator) is the experimental platform we developed which is able to evaluate the QoS performance of a city BSN. This is possible by acquiring real city topological data well as the relative PTS timetable. URBeS is able to accurately reproduce bus movements in real urban environments and to simulate data forwarding among buses. Our experimental platform is able to support any external routing policy in order to compare performances of various routing algorithms. The functional scheme of URBeS, reported in Fig. 2, aims to emphasize both the tri-modular design of its functionalities and its support for any externally specified routing policy.

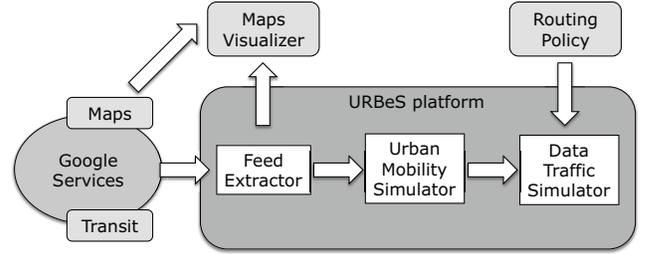

Fig. 2. URBeS functional scheme.

The analysis of a city PTS starts from a Google Transit [23] feed, which is a database of planned trips, provided by a transit authority from which we devise a movement model. GPS coordinates of every bus stop have later been converted to Cartesian coordinates. Nodes along each line move between coordinates accordingly to real timetables moving at a constant speed. Pauses at stops are – when planned – included in the transit feed and thus simulated accordingly.

The URBeS framework is then composed of three sequential modules.

In the first phase, URBeS parses information from the feed and produces a timetable of bus movements together with a topology of the PTS layout.

The output from the first phase is fed to an urban mobility simulation module which is in charge of generate mobility traces for all the buses based on the real PTS timetable. In this phase, URBeS also computes statistics about bus contacts. They are useful for predicting intra- and inter-contact times and for understanding city coverage of the PTS.

The last phase adds data traffic to the picture: network traffic is generated randomly by each bus and is delivered following the provided routing policy. URBeS logs detailed information to profile delivery rates, delays, and locations where forwarding takes place. Simulation is performed by means of a custom-made discrete event simulation software we developed around the concept of BSN. This development has been required in order to overcome poor scalability in term of total traffic and number of nodes from existing products (e.g., GloMoSim [24] and ns-2 [25]), while introducing an highly optimized urban canyon model which fails to be even in more modern simulation environments (e.g., the ONE simulator [26]). Results from our simulator have been positively evaluated against GloMoSim for low traffic levels using the same output coming from Urban Mobility Simulator. Our simulator can thus provide a valid comparison between different routing policies, while allowing easy testing on multiple urban environments.

In the following Sections we discuss the implementation details relative to the three components of URBeS.

### A. Google Transit Feed Extractor

The Google Transit Feed Extractor (FE) is actually a data converter tool which reads a feed and stores all data useful for simulation in a more manageable format.

A Google Transit feed is a database comprised by several tables reporting individual bus trips. Each trip is classified by a time context (e.g., Monday, Sunday, or "New Years Eve"), a route id, and a bus head sign. In order to build the topology, the FE starts from the list of trips and classifies all possible paths based on the stops made. Paths are then grouped together with reversals and aliases; reversals are paths making the same stops in reverse order, whereas aliases are routes having the same end of the line and almost the same intermediate stops (i.e., their difference is below a certain threshold). All paths in a group are identified as a single – looping – line, as traditionally understood by passengers. All lines must be closed paths as that way it is possible to assign buses for subsequent departures once they have completed a round trip.

Lines which does not define a closed path are not used for data transportation because it is not possible for us to make a reasonable assumptions about the future activity of buses running along them (such as in the case of shuttles making one-way trips from the train station to the airport where there is no timetable available for the way back). Thus, following the normal procedures, we would be required to consider a bus reaching the end of the line as to be going out of service and discarding all transported packets. Needless to say, such a behavior is not useful for routing where encounters are not deterministic and, as a consequence, such lines are not considered to be part of the data distribution system.

The output generated from the FE is a database of routes and timetables from real traces which are fed to the second module of URBeS.

### B. Urban Mobility Simulator

The Urban Mobility Simulator (UMS) generates mobility traces for all buses during a given period by using the database created by the FE.

The UMS module creates bus instances as needed to ensure scheduled starts and manages buses going out of services when the PTS is overpopulated. Buses leave the line head at the scheduled time and make all the stops according to the set timetable; while completing a path a bus waits at the head of the line, if necessary, for the next scheduled departure. If there are already two buses in line, then we consider it to be overpopulated and the bus goes out of service. In case of aliases, the bus waiting in line will select the path of the next aliased line departing from the current head of line.

Since location and timetables from Google Transit are driving nodes during simulation, UMS is completely decoupled from a movement model: we have deterministic information about system evolution and there is no need to add parameters coming from any external model. Give that in a real PTS, timetables are planned taking into account average daily traffic, along the day we are not required to predict and simulate traffic jams. Nevertheless, we added random noise of up to ten minutes on the scheduled departure times, in order to take into consideration small variations between average traffic and actual street condition. Moreover, to determine if two buses are in radio contact, a line-of-sight model is adopted. Urban canyons play a fundamental role in wireless links and we take them into account by using a street map created from bus routes.

At this stage, we compute a first analysis based on topology and bus movement. The analysis provides the evolution of bus populations and relative contacts, information about the number of neighbors, the and distribution of intra- and inter-contact times.

The output of the UMS is a trace of bus movements in a bi-dimensional space.

### C. Data Traffic Simulator

The Data Traffic Simulator (DTS) leverages the output of UMS by introducing network traffic and applying a routing policy.

When a packet is generated it is placed in the node local buffer until forwarding becomes possible; an isolated bus will keep accumulating packets when no contacts are experienced. When an encounter happens all the packets are checked for forwarding. During forwarding nodes are subject to limitations in term of transmission bandwidth (i.e., a limited number of packets per second in accordance with the available bandwidth) and buffer space. Forwarding can take place only if more packets can be sent in that second and buffer space is available at the destination. In case there are not enough resources to forward all packets a first-in-first-out policy is applied. Of course, while buffer space prevents forwarding during all the contact period bandwidth will still be available in the next second. When eligible, packets are forwarded in accordance with the adopted routing policy.

When a bus reaches the end of the line it may or may not queue up and wait for another scheduled departure. If the bus keeps a place in line it will hold all its data and will keep generating messages while waiting. If, on the other hand, the bus leaves service all content will be pushed to the first bus waiting in line. If there is no bus – because there are no more scheduled departures – all the stored messages are dropped and packets will be considered lost.

The routing policy used by the BSN is provided as an external module and may be either link-state or distance-vector. A user may implement his/her own routing policy by defining the application logic based on which forwarding will take place. This approach makes URBeS an ideal platform for routing protocol development and comparison: multiple experiments can be easily run specifying different routing modules to be used with the same traffic pattern.

The output of the DTS is a complete trace of the generated data traffic. From this trace we can compute all needed performance indexes and perform protocol analysis.

## IV. THE Op-HOP ROUTING ALGORITHM

Op-HOP (Opportunistic Hopping on Probabilities) is a novel probability-driven routing protocol for BSNs with the following features: it uses a probabilistic model based on the number of encounters rather than their durations, forwards packets using a single copy approach, and opportunistically exploits unplanned encounters to improve delivery performances. These encounter probabilities are computed by means of a field measurement aiming to estimate the actual reliability of a sequence of encounters rather than considering the intra-contact times duration as in [21]. Op-HOP algorithm looks for a convenient tradeoff between path lengths and their probabilities to occur: long paths are discarded in favor of shorter ones if their probabilities are not significantly higher. Finally, Op-HOP is not building a strict path for a packet to follow: it will start from a link-state approach but will also exploit opportunistic contacts occurring by chance to swerve data on a shorter or more reliable path. The use of a link-state approach is supported by the fact that PTS topology is not changing and timetables are known a priori.

The design guidelines for Op-HOP where:

1) *Generality.* Op-Hop relies only on data provided by a transit authority. Thus, it avoid using models for mobility and city traffic which may not be able to express the heterogeneity of real world cities.
2) *Robustness.* Routing performances must be independent of system perturbation: city topology, PTS design, reasonable traffic variations.
3) *Scalability.* The need to support great number of messages as well as the loose QoS requirements of the addressed applications lead to the use of a single copy forwarding strategy. This is to avoid uncontrolled traffic overhead, which could require additional resources.

Moreover, as already rationalized by [15], the routing scheme must work at the *bus-line level*. In other words, the next hop toward a certain destination concern (any bus of) a line, rather than a specific bus. This approach is in accordance with the observation that buses belonging to the same line have a high probability of encountering one another, both at the end of the line and when traveling in opposite directions; therefore, any message can easily get spread within the line.

### A. Routing on a Probabilistic Graph

The line overlaps of the PTS lead to the construction of a connectivity graph which represents all the possible routes that the messages can take. This problem can be represented as a graph $G = (V, E)$, in which each node $v \in V$ is a bus line and each edge $e \in E$ denotes the existence of at least one connection among buses belonging to the two lines (i.e., the end-point nodes). A path or route $r$ on $G$ represents the sequence of nodes (lines) a message has to traverse in order to be delivered. Being $G$ an unweighted graph, the optimal routing strategy is based on minimizing the number of hops; however, an important chip of information is missing. In this model, an edge informs that at least one connection exists among the buses of the two lines; nevertheless, how often the two lines "intersect" is not taken into account. In order to obtain this information a weighted graph can be used: the weight of an edge is a function of the actual probability of encounter between the two lines.

### B. Encounter Probability Model

Being $i$ and $j$ two bus lines, we denote with $p_{i,j}$ the probability that (any bus of) line $i$ encounters (any bus of) line $j$. For any bus $b$ traveling on line $i$, the following Bernoullian random variable $X_b(j)$ is defined:

$$X_b(j) = \begin{cases} 1 & \text{if bus } b \text{ encounters any bus of line } j \\ 0 & \text{otherwise} \end{cases} \quad (1)$$

During a day, a bus $b$ makes $t_b$ trips and collects a sample $x_b(j)$ of the random variable $X_b(j)$. At the end of the day the samples collected by all buses of line $i$ are aggregated, and the probability of encounter between line $i$ and $j$ is estimated as:

$$p_{i,j} = \frac{\sum_{b \in i} x_b(j)}{\sum_{b \in i} t_b} \quad (2)$$

Considering encounters between different pairs of lines as independent events, the overall probability of a route $r$, made of $d_r$ hops, is:

$$p_r = \prod_{k=1}^{d_r} p_{k,k+1} \quad (3)$$

where $k$ enumerates the nodes along the route $r$.

Given any pair of source and destination lines, $l$ and $m$, let $\mathcal{R}_{l,m}$ be the set of all routes $r_{l,m}$ connecting $l$ to $m$. The best path $r^*_{l,m}$ to deliver a packet from $l$ to $m$ is one in $\mathcal{R}_{l,m}$ maximizing the delivery probability:

$$r^*_{l,m} := \left\{ r^*_{l,m} : p_{r^*_{l,m}} \geq p_{r_{l,m}} \forall r_{l,m} \in \mathcal{R}_{l,m} \right\} \quad (4)$$

When applying this probabilistic model on real traces it is possible to create long path with high probabilities due to a large number of encounters of buses along the path. In such cases, due to the number of forwarding required, the high delivery probability is coupled with a potentially greater delivery delay. Despite delivery probability is the most important metric for us, it does not worth the effort to suffer a significant increase in term of traversed hops in order to raise this probability by a negligible amount. To constraint this phenomenon we decided to keep the probability model as Bernoullian, but to slightly change the probability of encouter estimates by truncating them to the first digit and reducing all probabilities of 1.0 to 0.9.

## C. Construction of the Weighted Graph

An edge between two nodes exists, on the graph $G$, if the probability of encounter between the two bus lines is greater than zero. To correctly assign weights to edges, we rely on the following chain of equalities:

$$\max_{r \in R} p_r = \max_{r \in R} \log \left( \prod_{k=1}^{d_r} p_{k,k+1} \right) \quad (5)$$

$$= \min_{r \in R} \sum_{k=1}^{d_r} \log \left( \frac{1}{p_{k,k+1}} \right) \quad (6)$$

Indeed, we define the weight $w_{i,j}$ of the edge connecting node $i$ and $j$ as follows:

$$w_{i,j} = \log \left( \frac{1}{p_{i,j}} \right) \quad (7)$$

To find the optimal routes, a shortest path algorithm – such as Dijkstra's – is applied to this probabilistically weighted directed graph $G = (V, E, W)$.

## D. Improving Routing Performance with Opportunistic Contacts

To make the proposed routing algorithm completely opportunistic, forwarding opportunities must be exploited. When a bus $b_1$ comes in contact with another bus $b_2$, the routing table is looked up. For every packet in its buffer, $b_1$ performs forwarding if either: (i) $b_2$ is the designated next hop, or (ii) $b_2$ has a shorter distance (or a better delivery probability) to the destination than the designated next hop.

## E. Delivery Delay Analysis

The average delivery time $T_{l,m}$ on a path $r$ between two nodes $l$ and $m$ can be estimated according to the following rationale. Assuming $i$ as any intermediate node between $l$ and $m$, let $T_i$ be the trip time needed for a bus on line $i$, and let $N_i$ be the number of trips before the encounter occurs between a bus on line $i$ and a bus part of the next hop line. Since the process has been modeled via a Bernoulli random variable, $N_i$ follows a geometric distribution whose mean value is given by $\frac{(1-p_{i,i+1})}{p_{i,i+1}}$. Moreover, also the time $T_i^0$, which is the time to wait from reception of the packet and the next chance to forward with the next hop line, must be taken into account. The overall delay is the sum of the times needed for every hop:

$$T_{l,m} = \sum_{i=1}^{v_r-1} \left( T_i^0 + N_i T_i \right) \quad (8)$$

where $v_r$ is the number of vertices on path $r$. Being $N_i$ and $T_i$ independent random variables and assuming that the average trip time $t_i$ is the one provided by the PTS authority, the expected value of the delay time is given by:

$$E[T_{l,m}] = \sum_{i=1}^{v_r-1} \left( \frac{t_i}{2} + \frac{(1-p_{i,i+1})}{p_{i,i+1}} t_i \right) \quad (9)$$

In our implementation we take into account opportunistic contacts so as to improve routing performance. Exploiting casual encounters allows us to identify Eq. 9 as the upper bound to the delivery delay in our routing algorithm.

## V. EVALUATION

In this section we evaluate our proposal using real PTS time schedules of three large cities. Here, we present benchmark results on three cities – Milan, Edmonton, and Chicago – as indicative of all the cities reviewed during the course of our research. A maps of the bus lines used in our simulations is shown in Fig. 3.

Initially, we identified the key features to understanding and predicting BSNs performance. We quantify these characteristics by measuring city size and bus line density, also adding topology as useful for the purposes of our project. Milan, Italy represents our first case study. It is a medium size town (typical for many European cities). The PTS of Milan is a complex system extending above and below ground. Despite this, due to the underground aquifer and archaeological remains, the subway system is quite underdeveloped, while the ground transportation system inside the city spans around 70 lines with a very high bus density. As it can be observed from Fig. 3(a), the overall city structure is clearly not manhattan-like; crosses between bus lines may occur at any time and there is no constant space between contacts. Moreover, buses can run along three kinds of routes: they may span across the city, run around the center, or cover only a peripheral section making a small loop. Example of such lines can be seen in Fig. 4(a), 4(b), and 4(c) respectively.

While manhattan-like topologies have been extensively studied in the past, uneven topologies like the one we just described are less addressed. This is Why Milan is our most interesting case study to test BSN routing: for its uneven topological structure and the frequent contacts between nodes.

The second city we consider is Edmonton, Alberta (Canada) whose PTS is depicted in Fig. 3(b). It is larger than Milan, has more bus lines (a total of 94) but with a lower density; the city map is mostly regular, with some winding streets in the suburbs. Differently from Milan, bus encounters are less frequent while in transit due to the lower density, nevertheless

The last city is Chicago, chosen expressly to test scalability due to its immense size. The PTS in this case can be seen in Fig. 3(c). Chicago has a great many bus lines (142), with a very low density, and a Manhattan-like plan.

Table I presents a summary of the three cities and relative PTS layouts. In the table, bus line density is calculated as the mean number of miles covered by buses per square mile.

In all of our experiments we used a radio range is 100 meters, as typical for IEEE 802.11b technology. Also, to take urban canyons into account, we consider only line-of-sight contacts.

In the following section, we analyze the PTS behavior and the expected performances by routing data over these three cities.

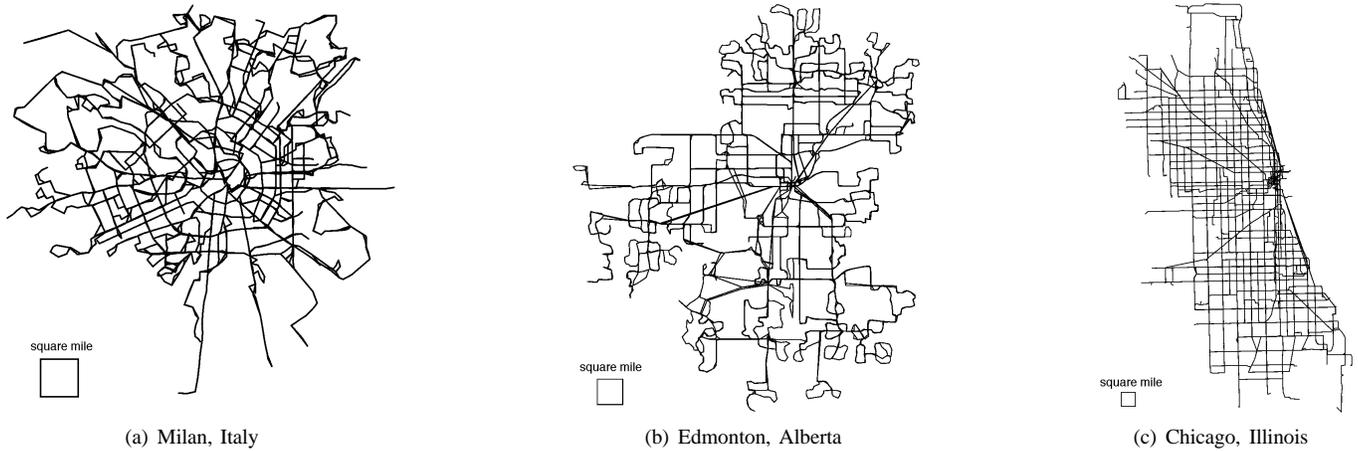

(a) Milan, Italy     (b) Edmonton, Alberta     (c) Chicago, Illinois

Fig. 3. Graphical picture of public transportation lines used in the analysis.

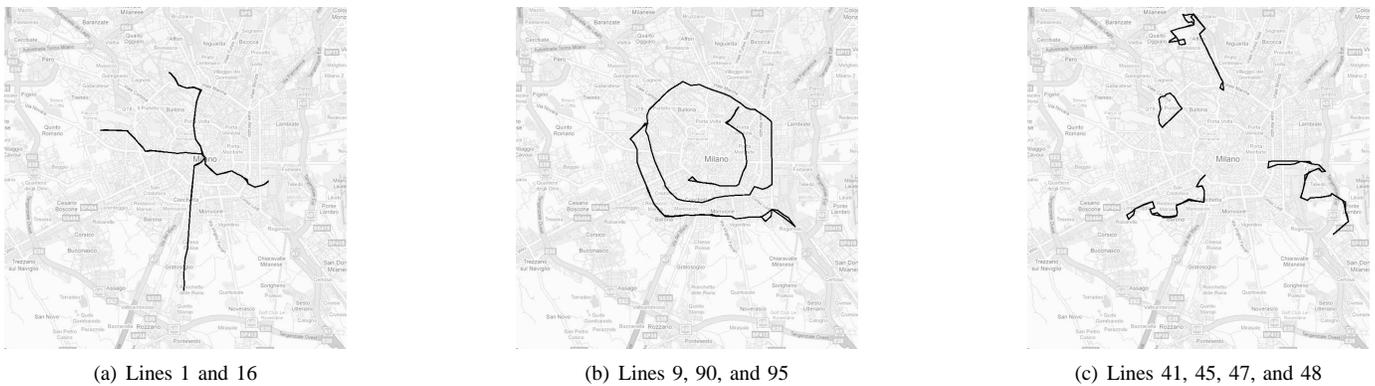

(a) Lines 1 and 16     (b) Lines 9, 90, and 95     (c) Lines 41, 45, 47, and 48

Fig. 4. Samples of different paths inside the city of Milan.

TABLE I
PROPERTIES OF PTS LAYOUTS FOR THE THREE CITIES.

|  | Milan | Edmonton | Chicago |
|---|---|---|---|
| City size in square miles | 49 | 134 | 380 |
| Number of lines | 69 | 94 | 142 |
| Line length (mean ± st. dev.) in miles | 9.9 ± 3.6 | 12.9 ± 6.7 | 15.3 ± 7.0 |
| Line density | 13.85 | 9.05 | 5.73 |

## A. PTS Analysis

We used the experimental platform introduced in Sec. III to simulate the behavior of the three PTSs. During simulations we consider only working days since the PTS scheduling is usually the same Monday to Friday. All simulations cover a full daily cycle of the public transportation service during a generic working day; from the moment when the first scheduled bus leaves up to the end of the last scheduled run.

Figure 5(a) reports the evolution over the course of a day for the bus population. Simulated bus population reproduces real world behavior: during the day we can observe two peaks associated to rush hours. The first rush hour happens when kids go to school and employees head for the office, while the second peak – which occurs as office workers commute back home – is not as high as in the morning because kids are already out of school. The number of buses is almost constant between rush hours and in the evening the population decreases to zero. The evolution of the number of contacts during the day follows the same trend as the population (Fig. 5(b)). The total number of contacts for Milan, Edmonton and Chicago are respectively $108,600$, $85,680$, and $166,722$.

Videos regarding bus movement simulations are available on the CARTOON project [27] website. Snapshots regarding buses position at 8 A.M. (when the number of buses is at maximum) is reported in Fig. 6.

In both the movies and pictures each dot represents a bus; a black dot is an isolated node while d different color represents a node which is in radio contact with at least one other node. After a transient, buses distributed evenly over the map;

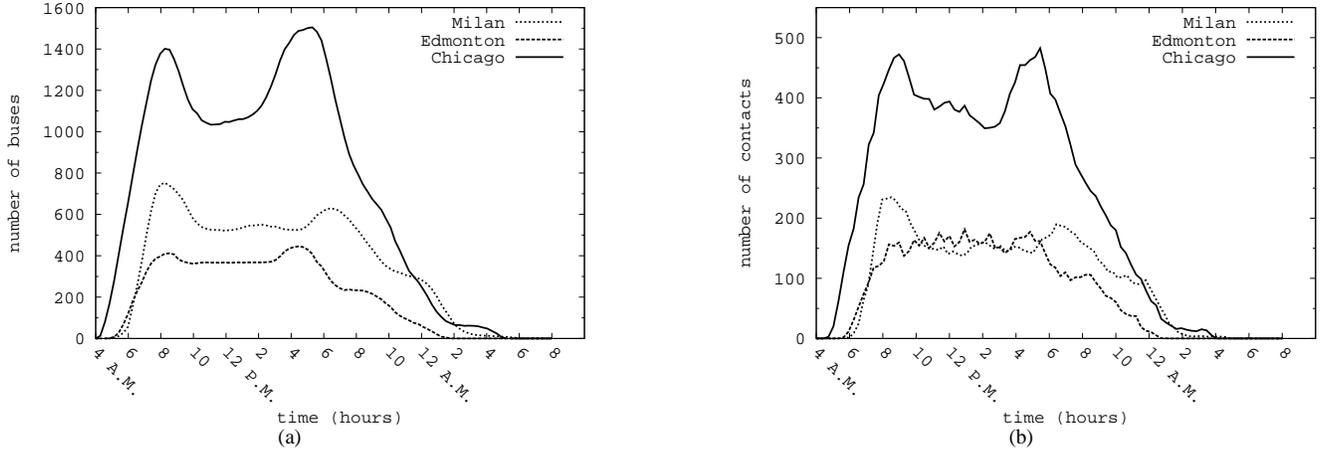

Fig. 5. Evolution of number of (a) buses and (b) contacts during the day.

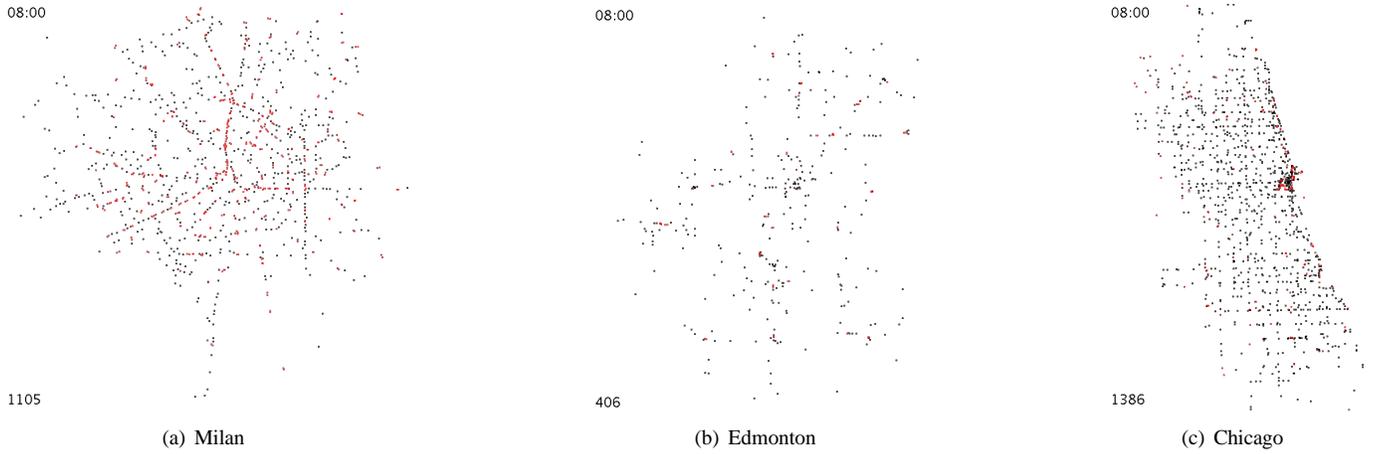

Fig. 6. Buses disposition for the three cities at 8 A.M., the total number of running buses is reported in each lower left corner.

in particular, color patches indicate the places with highest contact opportunities. We observe the three cities exhibit very different types of behaviors. In Chicago, connected nodes are concentrated mainly in the harbor area, along the coast, and at suburban ends of the lines. Edmonton seems to have a dozen main meeting points (shared ends of the line) and buses run from one meeting point to another, sometimes very quickly. In Milan, buses travel mostly along circular paths around the city center and on radial routes extending from downtown out to the suburbs. Consequently, we want to stress the fundamental differences from one city to the next which require a large-scale experimental analysis for purposes of evaluating the performance of routing algorithms.

Considerations about the PTS design may also be drawn from Fig. 7, where an histogram of the number of neighbors of all buses is reported. With the term "neighbors" we identify the number of buses that a bus is able to contact at a given time while running along its route. This figure confirms that not only the cities are different; rather different assumptions underpin the various PTSs. While Edmonton's is based on hubs where more lines converge and buses are rescheduled, Milan's aims for a more even distribution of the lines covering the city. Buses in Milan connect in pairs, while in Edmonton the distribution of the number of neighbors has a long tail on high values because of to the hubs. Similarly, the number of contacts in Edmonton is higher during the day due to the connection opportunities at the shared ends of the line.

TABLE II
INTRA-CONTACT TIMES (IN SECONDS).

|  | Milan | Edmonton | Chicago |
|---|---|---|---|
| Median | 20 | 22 | 14 |
| Mean | 38 | 261 | 132 |
| Std. dev. | 104 | 663 | 462 |

In the described scenarios, we also analyze both *intra-* and *inter-contact times*. Intra-contact times (see Tab. II) are defined as the duration of each contact between two buses. Intra-contact times manifest a low value for the median especially in Chicago; this result indicates a huge number of contacts at street intersections and leads to a low connection time, which implies a low encounter probability. On the contrary, mean values are high due to partially overlapping paths, as where they overlap on their routes buses may travel a significative distance together. Moreover, in Edmonton this latter factor is accentuated by the fact that the buses queue up for departure while data connectivity with many other lines is available.

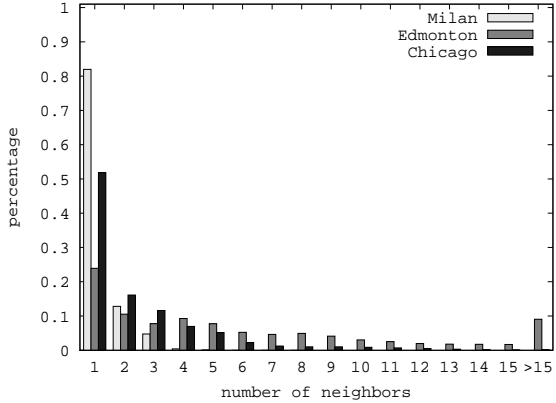

Fig. 7. Histogram of the number of neighbors.

TABLE III
INTER-CONTACT TIMES (IN SECONDS).

|  | Milan | Edmonton | Chicago |
|---|---|---|---|
| Median | 465 | 1118 | 453 |
| Mean | 2623 | 2647 | 3042 |
| Std. dev. | 4892 | 4138 | 5545 |

Inter-contact time for a node is defined as the time elapsed from the end of an encounter with a node part of line $l$ and the beginning of next encounter with any node belonging to $l$. Inter-contact times are reported in Tab. III and are very similar in all three scenarios.

### B. Analysis of Routing Protocols

An extensive analysis of real cities is required in order to prove the correctness of routing algorithms on BSNs. We compared different routing protocols, namely: minimum-hop routing, epidemic routing, Op-HOP, and the routing strategy proposed in [21] based on intra-contact times. The minimum-hop strategy, which tries to minimize the number of traversed lines, provides a benchmark comparison with the routing strategy used in wired networks. On the other hand, the epidemic routing is useful to identify a lower bound for delivery and to quantify resources saved by using a single-copy strategy.

To the best of our knowledge, the only previous proposal which employs a single-copy approach and has been tested on a real city (Shanghai) is the one proposed in [21]. For the sake of clarity, in the following this routing algorithm will be identified as "Shanghai". Shanghai applies the Dijkstra algorithm using intra-contact times as a metric, while in Op-HOP we propose the encounter probabilities.

The real differences between the considered algorithms lie in the number of hops of selected routing paths and in the probabilities in re that sequence of hops actually taking place.

The comparison of these routing algorithms, on all analyzed cities, produces similar results. Therefore, we describe only Edmonton in detail. For all possible paths in Edmonton the histograms of the number of hops are reported in Fig. 8(a), while the histograms of the encounter probabilities are presented in Fig. 8(b).

As can be noted, minimum hop provides many short paths characterized by a very low probability, whereas Shanghai raises the probabilities by increasing the number of hops. Op-HOP fills a spot in between: it maximizes the path probability to achieve a better delivery rate than Shanghai; at the same time, the path lengths are longer than the minimum hop case yet still bounded below Shanghai. Furthermore, this latter property leads to shorter delivery times as confirmed by simulations presented later in this section. In particular, Shanghai falls short of identifying a real contact opportunity as the total intra-contact times are not always proportional to the number of contacts.

Of course, the picture is not complete without a comparison with algorithms using a multicopy approach. We also compare our method with MaxProp [14] and Rapid [16] since they are the most widely adopted in literature. Of course, being multicopy, resource usage for these routing schemes is such that they work better on a small scale (e.g., in a campus) but simulations hinted that scalability on a complex urban scenario is limited.

### C. Performance Evaluation

We carried out an extensive number of simulations in the three aforementioned scenarios. In every well-organized PTS, timetables are planned taking into account average traffic during the course of the day. This feature helps us to give the simulations a good sense of realism with limited effort: we are not required to predict and simulate traffic jams. Nevertheless, we added some random noise in order to take into consideration small variations between average traffic and actual conditions on the street. In our simulations we perturbed each trip by adding a random noise of up to ten minutes w.r.t. the scheduled departure times.

We study the performances of the considered routing algorithms in terms of message delivery ratio and latency.

During simulations, the traffic is supposed to be generated from a specific bus to any bus belonging to a bus line, according to [15]. Data traffic generation is performed continuously during working hours, 8 a.m. to 6 p.m. Every five minutes each bus generates a message to be delivered to a randomly chosen bus line. The total number of packets coming into play in Milan, Edmonton, and Chicago are respectively $68,597$, $46,176$, and $147,851$. We used WiFi link technology, a bandwidth of 10 Mpbs as provided by 802.11b connectivity, and a packet size of 64 KB.

Here we evaluate the delivered packet percentages re the three cities, as shown in Fig. 9.

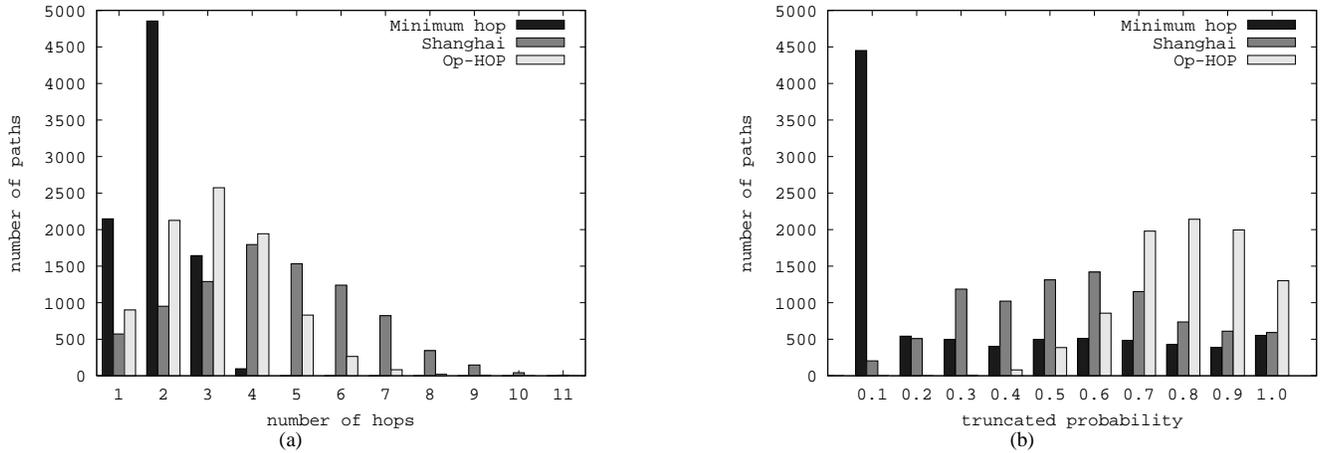

Fig. 8. Histograms of the number of (a) hops and (b) truncated path probabilities in Edmonton.

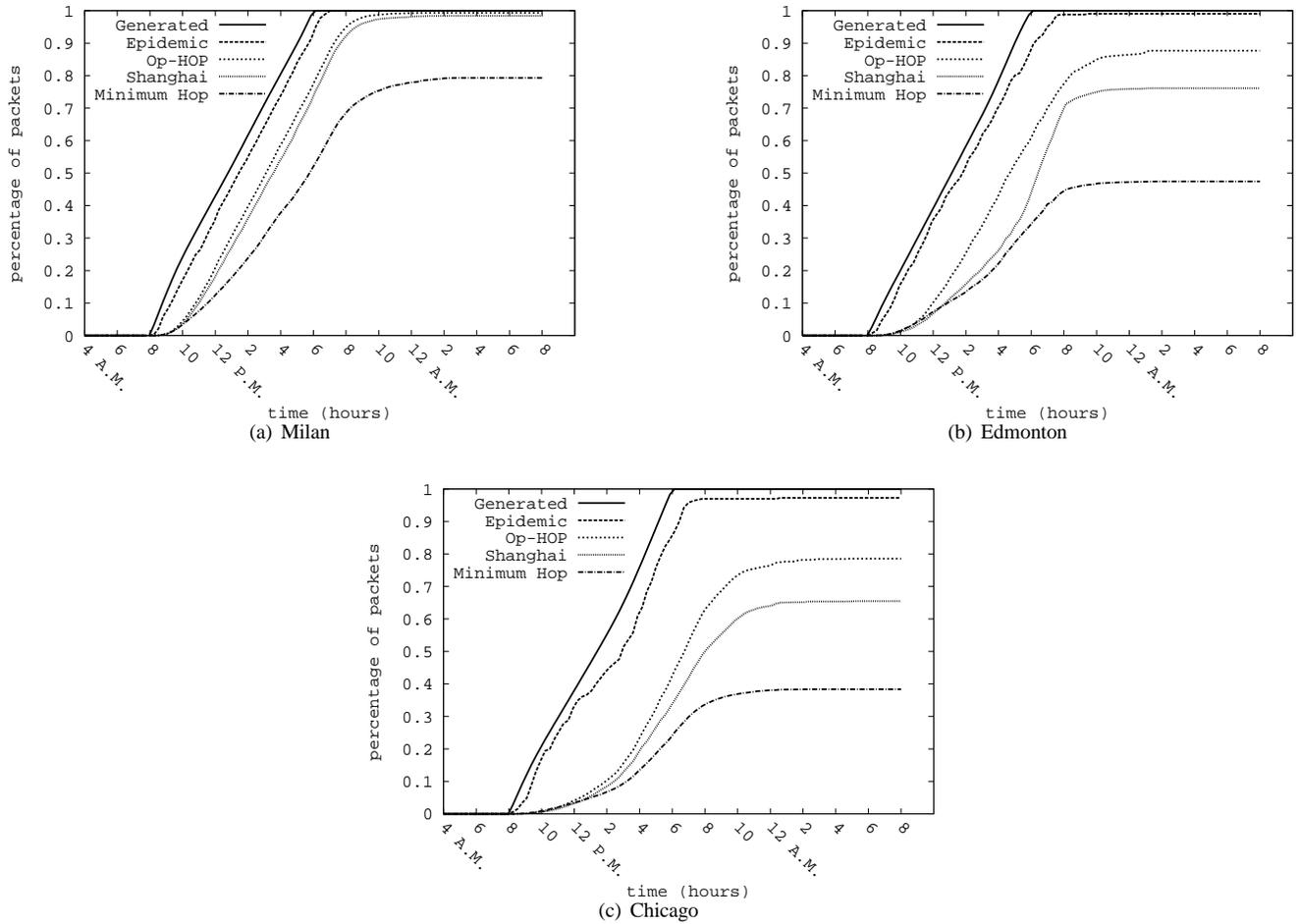

Fig. 9. Traffic generation and delivery given as a percentage of the total number of packets during the course of the day.

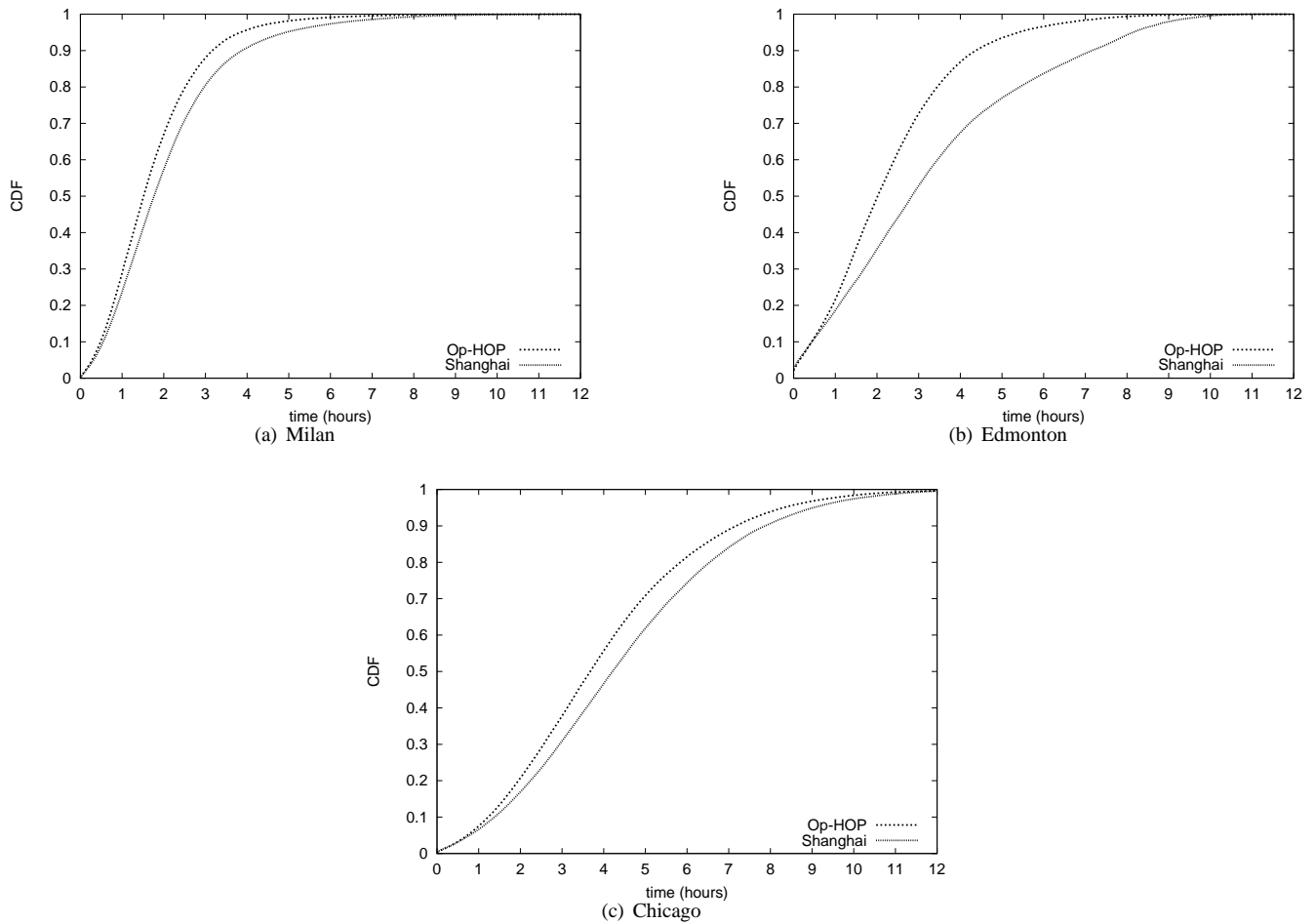

Fig. 10. Comparison of delivery delay cumulative distributions.

The minimum hop routing algorithm, which disregards property of encounters, performs very poorly in all cases. Therefore we cannot consider it a useful strategy.

By closely inspecting Edmonton in Fig. 9(b) and Chicago in Fig. 9(c), we observe that the epidemic routing as well is unable to reach a 100% delivery ratio. The packets might go undelivered when the bus population is low and the network gets heavily partitioned. Furthermore, if we use epidemic routing, the average number of packet replicas per message is $403$ in Milan, $283$ in Edmonton and $716$ in Chicago.

In view of the above considerations, minimum hop and epidemic routing strategies are clearly not suitable for PTS deployment in urban environments.

Let us now consider the Shanghai routing algorithm: it is always outperformed by Op-HOP, especially when the line density is low. In less connected environments like Edmonton and Chicago, the difference in performance is shown to be about 10% (see Fig. 9(b) and 9(c)).

The other key metric to evaluate protocols performance in DTN is delivery delay. Cumulative distribution of the delivery delay for the three cities is reported in Fig. 10.

As evident in the pictures, the delivery delay obtained by Op-HOP is always better than the one we get with Shanghai. In particular, in Edmonton, where intra-contact times are very long but cannot be associated to a high encounter probability along the path, a significant performance gap can be observed. For example, the two 0.9 quantiles differ by more than two and a half hours (Fig 10(b)).

In Tab. IV we present a summary of performances in re all cities and routing algorithms.

Overall, Op-HOP performances are always adequate for providing flexible urban-wide services, especially when the PTS is well engineered w.r.t. city topology. As a matter of fact, we reach a near 100% delivery rate in Milan and even in a huge and less interconnected city like Chicago we end up with a margin of undelivered packets of 20% or less. In terms of scalability and resources usage, moreover, Op-Hop greatly outperforms epidemic routing since it generates a single copy for every packet; and even in the worst of cases this advantage is only partially diminished by undeliveries.

The delivery delay is an important metric to quantify the accessibility to urban-wide services. From Tab. IV, we see that the end-to-end delay is limited to two hours in both large and average size cities. As for sprawling cities, we can reach

(a) Milan, Italy

|  | Epidemic | Min hop | Shanghai | Op-HOP |
|---|---|---|---|---|
| Delivery ratio | 100 % | 79.1 % | 98.3 % | **99.9 %** |
| Median time | 0.52 | 2.20 | 1.77 | **1.37** |
| Mean time | 0.54 | 3.12 | 2.11 | **1.61** |
| St. dev. time | 0.25 | 2.85 | 1.58 | **1.17** |

(b) Edmonton, Alberta

|  | Epidemic | Min hop | Shanghai | Op-HOP |
|---|---|---|---|---|
| Delivery ratio | 99.9 % | 45.9 % | 72.9 % | **87.7 %** |
| Median time | 0.93 | 1.51 | 2.84 | **1.83** |
| Mean time | 1.02 | 2.29 | 3.37 | **2.13** |
| St. dev. time | 0.69 | 2.30 | 1.52 | **1.55** |

(c) Chicago, Illinois

|  | Epidemic | Min hop | Shanghai | Op-HOP |
|---|---|---|---|---|
| Delivery ratio | 99.7 % | 37.7 % | 63.8 % | **82.6 %** |
| Median time | 1.13 | 2.92 | 4.21 | **2.86** |
| Mean time | 1.23 | 3.63 | 4.47 | **3.26** |
| St. dev. time | 0.79 | 2.81 | 2.51 | **2.14** |

TABLE IV
SUMMARY OF DELIVERY RATIOS AND TIMES (IN HOURS).

destinations in four hours circa, in line with the proposed applications.

By combining performance metrics as reported in Fig. 9 and 10, we obtain a diagram for expected QoS in every city (Fig. 11).

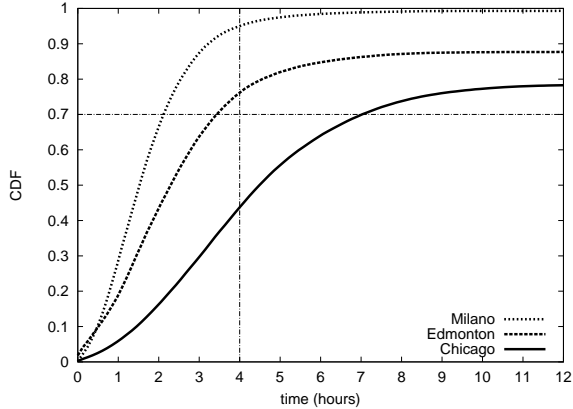

Fig. 11. Quality of Service diagram for Op-HOP.

Given the required QoS for the service we want to provide, from Fig. 11 we can grasp the percentage of the reachable population within a certain amount of time. As an example, following the vertical dash-dotted line we can see that within four hours Milan is able to deliver 0.97 of total traffic while Edmonton and Chicago 0.78 and 0.45 respectively. This figure also shows how long it takes for a message to reach a determinate share of destinations on BSNs. Similarly as before, following the horizontal dash-dotted line, we can see 0.70 of total traffic will be delivered within 2.1, 3.4, and 7.0 hours for Milan, Edmonton, and Chicago respectively.

Lastly, we conducted extensive experiments to evaluate the ability of the system to tolerate stressful traffic conditions. To this end, we let every bus generate from 1 to 60 packets per hour. Overall traffic depends on the number of buses; hence, the total amount of messages is huge, going all the way up to 739,167 packets a day in Chicago.

Figure 12(a) shows the worst case for buffer usage in such conditions. We can use these values to devise the on-board memory and to highlight possible bottlenecks. The buffer requirement is shown to increase linearly with the growing injected traffic, but even on extremely high workload the maximum value is bounded to a technically acceptable figure. Nowadays in fact, 0.5/1 GB of memory is available on any type of small and portable device.

In Fig. 12(b) and 12(c) we see that both delivery ratio and delay are mostly constant on load, as also while a loss of performance shows up only when the total traffic is very high and much influenced by the citys PTS density.

Nonetheless, the observed data allow us to state that BSN performances still ensure the envisioned QoS requirements even when traffic is intense – no matter the city.

As a last note, regarding scalability when comparing our proposed solution with multicopy approaches, simulation results for the city of Milan are presented in Fig. 13. In this figure performance indexes are evaluated when increasing the traffic generated by each bus in the system from 20 up to 80 packets per hour. In Fig. 13(a) we can clearly see that average delivery delay is in fact increasing for MaxProp and Rapid with the offered load. This phenomenon is due to the rising number of copies required for packets forwarding which are clogging the resources (both buffers and bandwidth), preventing an effective delivery during contacts. In Fig. 13(b) we observe a better situation since the packet loss increase is quite limited (albeit present) and Op-HOP is able to deliver more effectively only when the offered load id greater than 60 packets per hour generated by every running bus.

## VI. CONCLUSIONS

We have identified a large class of mobile applications whose requirements can be hardly addressed by the combined use of cellular networks and the current Internet. On the contrary, we show that Bus Switched Networks are a ready to use and quickly deployable platform able to satisfy all the envisioned networking and Quality of Service needs.

This paper presents a novel probabilistic routing protocol that leverages the bus encounters to achieve delivery efficiency while controlling resource consumption. We evaluate our proposal through an extensive benchmark analysis on three real cities selected so as to explore geo and structural diversity. Finally, a novel experimental platform (URBeS) is made available to facilitate the engineering of any project that aims effectively to deploy a BSN in one of the worlds cities.

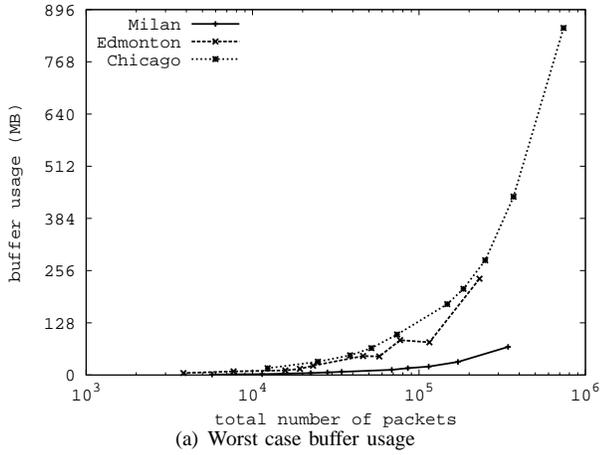
(a) Worst case buffer usage

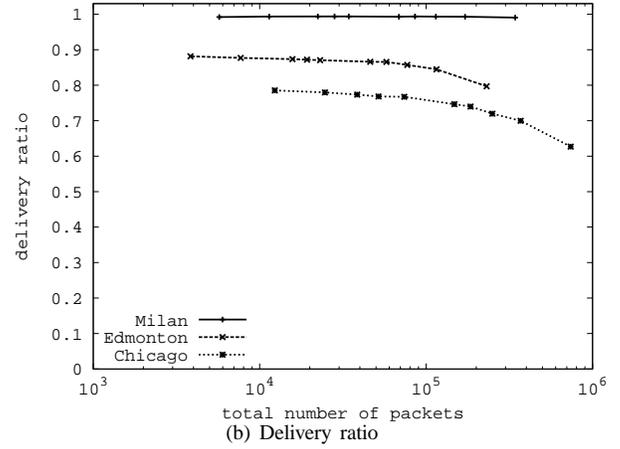
(b) Delivery ratio

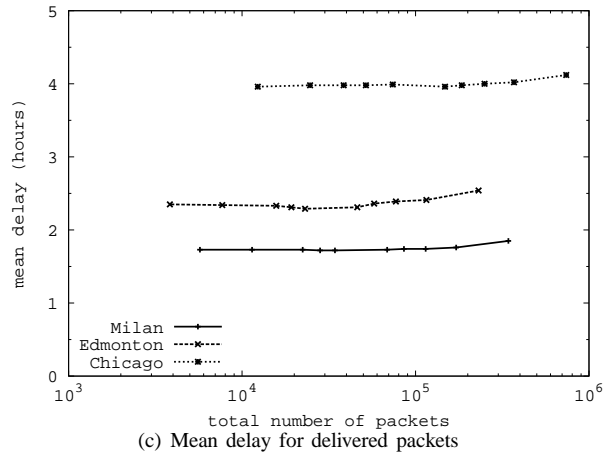
(c) Mean delay for delivered packets

Fig. 12. Evaluation of BSNs under different traffic load conditions.

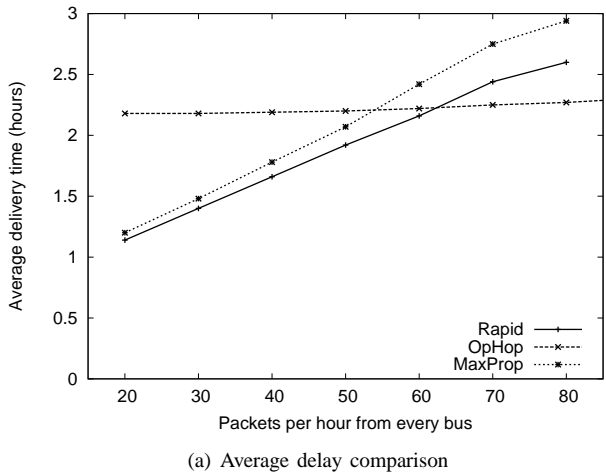
(a) Average delay comparison

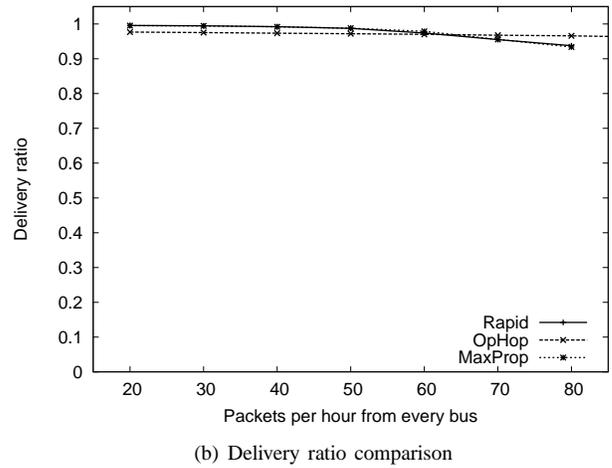
(b) Delivery ratio comparison

Fig. 13. Scalability comparison between Op-HOP, MaxProp, and Rapid on the city of Milan.